\newcommand{\CLASSINPUTtoptextmargin}{0.8in}
\newcommand{\CLASSINPUTbottomtextmargin}{1.03in}
\documentclass[conference,a4paper]{IEEEtran}
\IEEEoverridecommandlockouts
\usepackage{cite}
\usepackage{amsmath,amssymb,amsfonts}
\usepackage{algorithmic}
\usepackage{graphicx}
\usepackage{textcomp}
\usepackage{xcolor}
\usepackage{pgfplots}
\usepackage{tikz}
\usepackage{subcaption} %
\usepackage{mathtools} %
\usepackage{comment}    %
\usetikzlibrary{fit}
\usetikzlibrary{calc, shapes.gates.logic.US, shapes.gates.logic.IEC}
\usetikzlibrary{positioning}
\usetikzlibrary{spy,backgrounds}
\usetikzlibrary{shapes.multipart}
\usepackage{multirow}  
\usepackage{placeins} %
\usepackage{soul}   %
\pgfplotsset{compat=1.16}
\definecolor{darkblue}{rgb}{0,0.2706,0.541}
\definecolor{darkgreen}{rgb}{0.098,0.4784,0.5176}
\definecolor{lightgreen}{rgb}{0.3412,0.7411,0.7647}
\definecolor{orange}{rgb}{0.9255,0.4313,0}
\definecolor{darkred}{rgb}{0.7529,0,0}
\definecolor{black}{rgb}{0,0,0}
\def\BibTeX{{\rm B\kern-.05em{\sc i\kern-.025em b}\kern-.08em
    T\kern-.1667em\lower.7ex\hbox{E}\kern-.125emX}}

\makeatletter
\def\ps@IEEEtitlepagestyle{%
	\def\@oddfoot{\mycopyrightnotice}%
	\def\@evenfoot{}%
}
\def\mycopyrightnotice{%
	{\footnotesize
		This work has been submitted to the IEEE for possible publication. Copyright may be transferred without notice, after which this version may no longer be accessible.\hfill}%
	\gdef\mycopyrightnotice{}%
}
\makeatother
\begin{document}

\title{%
Implementation-Efficient \\ Finite Alphabet Decoding of Polar Codes
}

\author{\IEEEauthorblockN{Philipp Mohr, Syed Aizaz Ali Shah and Gerhard Bauch}
\IEEEauthorblockA{\textit{Institute of Communications} \\
\textit{Hamburg University of Technology}\\
Hamburg, Germany \\
\{philipp.mohr; aizaz.shah; bauch\}@tuhh.de}
}

\maketitle

\begin{abstract}
An implementation-efficient finite alphabet decoder for polar codes relying on coarsely quantized messages and low-complexity operations is proposed. Typically, finite alphabet decoding performs concatenated compression operations on the received channel messages to aggregate compact reliability information for error correction. These compression operations or mappings can be considered as lookup tables. For polar codes, the finite alphabet decoder design boils down to constructing lookup tables for the upper and lower branches of the building blocks within the code structure. A key challenge is to realize a hardware-friendly implementation of the lookup tables. 
This work uses the min-sum implementation for the upper branch lookup table and, as a novelty, a computational domain implementation for the lower branch lookup table.
The computational domain approach drastically reduces the number of implementation parameters. Furthermore, a restriction to uniform quantization in the lower branch allows a very hardware-friendly compression via clipping and bit-shifting. Its behavior is close to the optimal non-uniform quantization, whose implementation would require multiple high-resolution threshold comparisons.
Simulation results confirm excellent performance for the developed decoder. Unlike conventional fixed-point decoders, the proposed method involves an offline design that explicitly maximizes the preserved mutual information under coarse quantization. 
\end{abstract}

\section{Introduction}
Polar codes are the first class of linear block codes that have been shown to asymptotically achieve the capacity of binary-input discrete memory-less channels through successive cancellation (SC) decoding\cite{arikan_channel_2009}. While the SC decoding does not reach capacity for practical code word lengths, the introduction of successive cancellation list (SCL) decoding with cyclic redundancy check (CRC) \cite{tal_list_2015} made polar codes very competitive in the short-block length regime. Further advances eventually evolved polar codes to be standardized for the uplink and downlink control channels in 5G\cite{3gpp_38_212}, making them widely used nowadays. 

In a communication system, forward error correction requires a high proportion of the total energy and hardware resources for the baseband processing. In particular, the bit-width $w$ of the messages, which represent reliability information in the decoding process, should be chosen as small as possible to achieve the required performance with minimal space complexity. This led to the paradigm of finite alphabet decoding where $w$-bit integer-valued messages communicate reliability levels among lower and upper branch operations in the decoding graph of a polar code. Inherently, each multiple-input operation must involve a compression to maintain small bit widths in the output messages.

Recently, the information bottleneck (IB) method has been introduced for designing mutual information maximizing decoding operations implemented via lookup tables\cite{shah_design_2019,shah_coarsely_2019,shah_MSIB_2023,Akino_IB_Polar_japan_19}. However for a size $N$ code, $2N-2$ individual lookup tables with size of up to $2^{2w+1}$ are required\cite{shah_coarsely_2019}.

In this paper, we propose to use another implementation variant for decoding polar codes by using a so-called computational domain that avoids the multi-input lookup tables. The technique is inspired by the computational domain used in mutual information maximizing decoding of low-density parity check (LDPC) codes\cite{he_mutual_2019}. For each update, two messages are translated to representation levels and merged using an addition whose result is compressed via threshold comparisons. When symmetric representation levels are enforced, the number of implementation parameters is reduced to $2^{w}{+}2^{w{-}1}{-}1$ for the two translations and the compression. It can be shown that non-uniformly placed thresholds can preserve the same amount of mutual information as the lookup table approach~\cite{he_mutual_2019}. 

In \cite{mohr_uniform_2022} a simplified computational domain approach for LDPC decoding was proposed. We adopt the idea in the lower branch update of a polar decoder: A restriction to uniformly placed thresholds is exploited in order to effectively avoid the threshold comparisons, reducing the number of calculation operations from $w$ to $1$ for each of the $N_L\frac{1}{2}N\log_2(N)$ lower branch updates where $N_L$ is the list size. The uniform quantization is done with a very simple clipping and bit-shifting operation combined with properly scaled translated messages.
For the upper branch, the idea from \cite{shah_MSIB_2023} is kept, i.e., 
 the upper branch updates are designed using the min-sum rule.
The overall result is a highly implementation-efficient finite alphabet decoder, specified by only $2^w{+}1$ instead of $2^{2w{+}1}$ parameters per lower branch update compared to an IB decoder \cite{shah_design_2019,shah_coarsely_2019}.
The contributions can be summarized as follows:
\begin{itemize}
    \item A computational domain approach, known from LDPC decoding, is adopted for the decoding of polar codes. Its behavior is potentially equivalent to a mutual-information-maximizing lookup table based IB decoder but reduces the number of design parameters drastically.
    \item A simplified computational domain update is proposed that avoids costly threshold comparisons in each lower branch update at close-to-optimal performance. %
    \item Simulation results confirm that the proposed simplified decoder involves a loss of only 0.04-0.07\,dB for code rates ranging between 0.75 to 0.25.
\end{itemize}
The rest of the paper is organized as follows:
First section \ref{sec:polarcodes} briefly explains polar codes and their conventional LLR-based decoding. Then, section \ref{sec:finitealphabetdecoders} introduces the principle of finite alphabet decoding. In section \ref{sec:proposeddecoder} a new finite alphabet decoder variant with low complexity is described. Finally, section \ref{sec:performanceevaluation} evaluates the performance with block error rate simulations.

\section{Polar Codes}\label{sec:polarcodes}
A polar code with length $N=2^n$, where $n=1,2,\ldots$, is described by its $N\times N$ generator matrix 
\begin{equation}
\mathbf{G} = {\mathbf{F} }^{\otimes n}\mathbf{B},
\label{eq:NxN_generator_matrix}
\end{equation} 
where matrix $\mathbf{F}{=}\begin{bmatrix}
1	&	1 \\
0	&	1
\end{bmatrix} $ and $\mathbf{B}$ is the \textit{bit reversal} permutation matrix\cite{arikan_channel_2009}.
For a code rate of $R{=}K/N$, $N{-}K$ bits in $\textbf{u}=[u_0, \dots u_{N-1}]^T$ are set to fixed values, e.g., $u_i {=} 0$, and referred to as the \textit{frozen} bits. The values and locations of the frozen bits are known to the decoder. The remaining  $K$ positions, specified in the \textit{information} set $\mathcal{A}$, in \textbf{u} carry the information bits.
The process of determining the information set is referred to as the code construction.
The encoding  follows as $\textbf{x} {=} \mathbf{G} \mathbf{u}$.

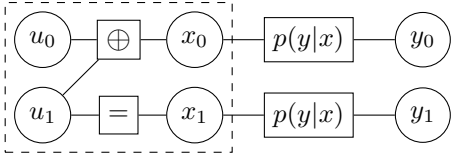
\begin{figure}[tbp]
	\centering      
	\begin{tikzpicture}[yscale=0.4, xscale=1, node distance=0.3cm, auto]
	
		\def \nodesize {0.5} %
		\def \VertDist {2.5}
		\def \HorDist {1}
		\def \HorDisplace {0.5}
		
		\node (U0) at (0-\HorDist,0) [draw, circle, minimum width = \nodesize cm]{$u_0$};
		\node (U1) at (0-\HorDist,0-\VertDist) [draw, circle, minimum width = \nodesize cm]{$u_1$};%
		\node (Cn0) at (0,0) [draw, rectangle, minimum width = \nodesize cm,minimum height = \nodesize cm]{\large$\oplus$};
		\node (Cn1) at (0,0-\VertDist) [draw, rectangle, minimum width = \nodesize cm,minimum height = \nodesize cm]{$=$};	
		\node (X0) at (\HorDist,0) [draw, circle, minimum width = \nodesize cm]{$x_0$};
		\node (X1) at (\HorDist,0-\VertDist) [draw, circle, minimum width = \nodesize cm]{$x_1$};
		\node (Cn2) at (2*\HorDist +\HorDisplace,0) [draw, rectangle, minimum width = \nodesize cm,minimum height = \nodesize cm]{$p(y|x)$};
		\node (Cn3) at (2*\HorDist +\HorDisplace,0-\VertDist) [draw, rectangle, minimum width = \nodesize cm,minimum height = \nodesize cm]{$p(y|x)$};	
		\node (Y0) at (3*\HorDist +2*\HorDisplace,0) [draw, circle, minimum width = \nodesize cm]{$y_0$};
		\node (Y1) at (3*\HorDist +2*\HorDisplace,0-\VertDist) [draw, circle, minimum width = \nodesize cm]{$y_1$};

		\draw[-] (U0) -- (Cn0);
		\draw[-] (U1) -- (Cn1);
		\draw[-] (Cn0) -- (X0);
		\draw[-] (Cn1) -- (X1);
		\draw[-] (U1) -- (Cn0);
		\draw[-] (X0) -- (Cn2) --(Y0);
		\draw[-] (X1) -- (Cn3) --(Y1);
		
		\node[draw,dashed,fit=(U0) (X1)] {};
	
	\end{tikzpicture}	
    \caption{ Factor graph of the building block (dashed rectangle) of a polar code and the transmission channel.}
    \label{fig:Tannergraph2.1}
   \vskip -10pt
\end{figure}

The matrix $\mathbf{F}$, as depicted in the factor graph of Fig. \ref{fig:Tannergraph2.1}, serves as the building block of polar codes.
It encodes  the bits $\mathbf{u}{=}[u_0,u_1]^T$ into the codeword  $\mathbf{x}{=}[x_0,x_1]^T$ which is transmitted over a channel with transition probabilities $p(y_i|x_i)$. The received codeword is $\mathbf{y}{=}[y_0,y_1]^T$. Two virtual bit channels are created over the building block:
The first bit channel treats $u_0$ as input and $\mathbf{y}_0 = \mathbf{y}$ as output, where $u_1$ is considered a hidden variable observed via $y_1$.
The second bit channel treats $u_1$ as input and  $\mathbf{y}_1 = \left[ \mathbf{y},u_0\right]^T$ as output, assuming the true knowledge of $u_0$.

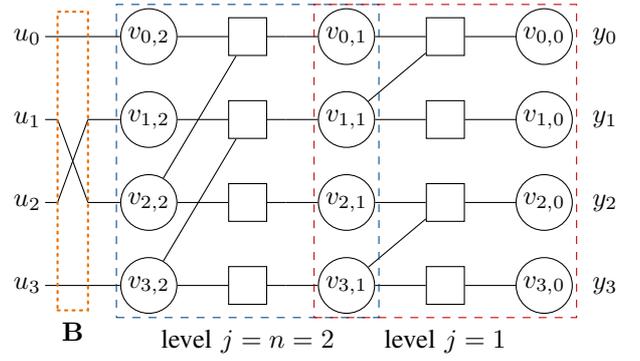
\begin{figure}[tbp]
	\centering
	\begin{tikzpicture}[yscale=0.5, xscale=1, node distance=0.3cm, inner sep = 0.5mm]

\definecolor{darkblue}{rgb}{0,0.2706,0.541}
\definecolor{darkred}{rgb}{0.7529,0,0}

\def \nodesize {0.5} %
\def \VertDist {2.2}
\def \HorDist {1.3}
\def \distRed {0.8} %

\def \XRefpoint {-4}
\def \YRefpoint {0}
\node (U02) at (\XRefpoint-\HorDist,\YRefpoint) [draw, circle, minimum width = \nodesize cm]{$v_{0,2}$};
\node (U12) at (\XRefpoint-\HorDist,\YRefpoint-\VertDist) [draw, circle, minimum width = \nodesize cm]{$v_{1,2}$};
\node (Cn0) at (\XRefpoint,\YRefpoint) [draw, rectangle, minimum width = \nodesize cm,minimum height = \nodesize cm]{};
\node (Cn1) at (\XRefpoint,\YRefpoint-\VertDist) [draw, rectangle, minimum width = \nodesize cm,minimum height = \nodesize cm]{};	
\coordinate (V0) at (\XRefpoint+\HorDist - \distRed,\YRefpoint);
\coordinate (V1) at (\XRefpoint+\HorDist - \distRed,\YRefpoint-\VertDist);
\draw[-] (U02) -- (Cn0);
\draw[-] (U12) -- (Cn1);
\draw[-] (Cn0) -- (V0);
\draw[-] (Cn1) -- (V1);

\node (u0) at (\XRefpoint-\HorDist- \distRed*2,\YRefpoint) []{$u_0$};
\node (u1) at (\XRefpoint-\HorDist- \distRed*2,\YRefpoint-\VertDist) []{$u_1$};

\def \XRefpoint {-4}
\def \YRefpoint {-4.4}
\node (U22) at (\XRefpoint-\HorDist,\YRefpoint) [draw, circle, minimum width = \nodesize cm]{$v_{2,2}$};
\node (U32) at (\XRefpoint-\HorDist,\YRefpoint-\VertDist) [draw, circle, minimum width = \nodesize cm]{$v_{3,2}$};
\node (Cn2) at (\XRefpoint,\YRefpoint) [draw, rectangle, minimum width = \nodesize cm,minimum height = \nodesize cm]{};
\node (Cn3) at (\XRefpoint,\YRefpoint-\VertDist) [draw, rectangle, minimum width = \nodesize cm,minimum height = \nodesize cm]{};	
\coordinate (V2) at (\XRefpoint+\HorDist - \distRed,\YRefpoint);
\coordinate (V3) at (\XRefpoint+\HorDist - \distRed,\YRefpoint-\VertDist);
\draw[-] (U22) -- (Cn2);
\draw[-] (U32) -- (Cn3);
\draw[-] (Cn2) -- (V2);
\draw[-] (Cn3) -- (V3);
\draw[-] (U22) -- (Cn0);
\draw[-] (U32) -- (Cn1);

\node (u2) at (\XRefpoint-\HorDist- \distRed*2,\YRefpoint) []{$u_2$};
\node (u3) at (\XRefpoint-\HorDist- \distRed*2,\YRefpoint-\VertDist) []{$u_3$};

\def \XRefpoint {-1.4}
\def \YRefpoint {0}
\node (U01) at (\XRefpoint-\HorDist,\YRefpoint) [draw, circle, minimum width = \nodesize cm]{$v_{0,1}$};
\node (U11) at (\XRefpoint-\HorDist,\YRefpoint-\VertDist) [draw, circle, minimum width = \nodesize cm]{$v_{1,1}$};%
\node (Cn0) at (\XRefpoint,\YRefpoint) [draw, rectangle, minimum width = \nodesize cm,minimum height = \nodesize cm]{};
\node (Cn1) at (\XRefpoint,\YRefpoint-\VertDist) [draw, rectangle, minimum width = \nodesize cm,minimum height = \nodesize cm]{};	
\node (X0) at (\XRefpoint+\HorDist,\YRefpoint) [draw, circle, minimum width = \nodesize cm]{$v_{0,0}$};
\node (X1) at (\XRefpoint+\HorDist,\YRefpoint-\VertDist) [draw, circle, minimum width = \nodesize cm]{$v_{1,0}$};
\draw[-] (U01) -- (Cn0);
\draw[-] (U11) -- (Cn1);
\draw[-] (Cn0) -- (X0);
\draw[-] (Cn1) -- (X1);
\draw[-] (U11) -- (Cn0);

\node (y0) at (\XRefpoint+\HorDist+ \distRed,\YRefpoint) []{$y_0$};
\node (y1) at (\XRefpoint+\HorDist+ \distRed,\YRefpoint-\VertDist) []{$y_1$};

\def \YRefpoint {-4.4}
\node (U21) at (\XRefpoint-\HorDist,\YRefpoint) [draw, circle, minimum width = \nodesize cm]{$v_{2,1}$};
\node (U31) at (\XRefpoint-\HorDist,\YRefpoint-\VertDist) [draw, circle, minimum width = \nodesize cm]{$v_{3,1}$};
\node (Cn2) at (\XRefpoint,\YRefpoint) [draw, rectangle, minimum width = \nodesize cm,minimum height = \nodesize cm]{};
\node (Cn3) at (\XRefpoint,\YRefpoint-\VertDist) [draw, rectangle, minimum width = \nodesize cm,minimum height = \nodesize cm]{};	
\node (X2) at (\XRefpoint+\HorDist,\YRefpoint) [draw, circle, minimum width = \nodesize cm]{$v_{2,0}$};
\node (X3) at (\XRefpoint+\HorDist,\YRefpoint-\VertDist) [draw, circle, minimum width = \nodesize cm]{$v_{3,0}$};
\draw[-] (U21) -- (Cn2);
\draw[-] (U31) -- (Cn3);
\draw[-] (Cn2) -- (X2);
\draw[-] (Cn3) -- (X3);
\draw[-] (U31) -- (Cn2);

\node (y2) at (\XRefpoint+\HorDist+ \distRed,\YRefpoint) []{$y_2$};
\node (y3) at (\XRefpoint+\HorDist+ \distRed,\YRefpoint-\VertDist) []{$y_3$};

\draw[-] (V0) -- (U01);
\draw[-] (V1) -- (U11.west);
\draw[-] (V2) -- (U21.west);
\draw[-] (V3) -- (U31);
\def \XRefpoint {-4}
\def \YRefpoint {0}
\coordinate (L0) at (\XRefpoint-\HorDist - \distRed*1.5,\YRefpoint);
\coordinate (L1) at (\XRefpoint-\HorDist - \distRed*1.5,\YRefpoint-\VertDist);
\coordinate (L2) at (\XRefpoint-\HorDist - \distRed*1.5,\YRefpoint-\VertDist*2);
\coordinate (L3) at (\XRefpoint-\HorDist - \distRed*1.5,\YRefpoint-\VertDist*3);
\coordinate (R0) at (\XRefpoint-\HorDist - \distRed*1,\YRefpoint);
\coordinate (R1) at (\XRefpoint-\HorDist - \distRed*1,\YRefpoint-\VertDist);
\coordinate (R2) at (\XRefpoint-\HorDist - \distRed*1,\YRefpoint-\VertDist*2);
\coordinate (R3) at (\XRefpoint-\HorDist - \distRed*1,\YRefpoint-\VertDist*3);

\draw[-] (L0) -- (R0);	\draw[-] (R0) -- (U02);
\draw[-] (L1) -- (R2);	\draw[-] (R1) -- (U12);
\draw[-] (L2) -- (R1);	\draw[-] (R2) -- (U22);
\draw[-] (L3) -- (R3);	\draw[-] (R3) -- (U32);

\draw[-] (L0) -- (u0);
\draw[-] (L1) -- (u1);
\draw[-] (L2) -- (u2);
\draw[-] (L3) -- (u3);

\coordinate (L0) at (\XRefpoint-\HorDist - \distRed*1.5,\YRefpoint+\VertDist*0.3);
\coordinate (L3) at (\XRefpoint-\HorDist - \distRed*1.5,\YRefpoint-\VertDist*3-\VertDist*0.3);
\coordinate (R0) at (\XRefpoint-\HorDist - \distRed*1,\YRefpoint+\VertDist*0.3);
\coordinate (R3) at (\XRefpoint-\HorDist - \distRed*1,\YRefpoint-\VertDist*3-\VertDist*0.3);
\node[fit={(L0) (R3)}, inner sep=0pt, draw=blue,dotted, color=orange,line cap=round,thick] (BR) {};
\node[below, inner sep = 4pt] at (BR.south) {$\mathbf{B}$ };

\node[draw,dashed,color=darkblue,fit=(U02) (U31)] (l2) {};
\node[draw,dashed,color=darkred,fit=(U01) (X3)] (l1) {};

\node[below, inner sep = 4pt] at (l2.south) {level $j=n=2$ };
\node[below, inner sep = 4pt] at (l1.south) {level $j=1$};
\end{tikzpicture}
	\caption{Structure of a polar code with length $N=4$. 
 }
 \vspace{-0.5cm}
	\label{fig:Tannergraph4}	
\end{figure}

The individual bit channels of a polar code of length $N{>}2$ are synthesized using a recursive application of the building block \cite{arikan_channel_2009}.
For instance, 
Fig. \ref{fig:Tannergraph4} shows the factor graph of a polar code for $N{=}4$, where the transmission channel is implicitly included, i.e., the right most variable nodes correspond to the (quantized) channel outputs $y_i$.
The code structure is composed of columns of building blocks referred to as  \textit{levels}, $j$ (dashed rectangles). Every node is labelled $v_{i,j}$ with row indices $i{=}0,1,\dots,N{-}1$, referred to as   \textit{stage}, and column indices $j{=}0,\dots, n$.
Then $v_{i,n}$ are the encoder inputs $u_i$ and $v_{i,0}$ are the channel outputs $y_i$.
From encoding perspective, Fig. \ref{fig:Tannergraph4} shows the flow of bits from left to right. From decoding perspective, LLRs flow from right to left in the code structure. 
of Fig. \ref{fig:Tannergraph4}.

\subsection{Successive Cancellation Decoding}\label{subsec:SC}
The successive cancellation decoder\cite{arikan_channel_2009} exploits the bit channels created in the code structure.
For a codeword length $N$, the SC decoder estimates the input of the $i$th bit channel, i.e., $\hat{u}_i$, in a sequential manner from $i=0$ to $N-1$.
With the bit channel output $\mathbf{y}_i = \left[\mathbf{y},u_0,\dots,u_{i-1} \right]^T$, $\hat{u}_i$ is estimated  at each  decoding stage $i\in \mathcal{A}$ as
\begin{equation}\label{eq:SC_decoding_rule}
    \hat{u}_i= 
        \begin{cases}
        0	&	L_{u_i}(\mathbf{y}_i) \geq 0 	\\
        1	&	\text{otherwise}.
        \end{cases}		
\end{equation}
The LLR $L_{u_i}(\mathbf{y}_i) = \log \frac{p(\mathbf{y}_i|u_i = 0)}{p(\mathbf{y}_i|u_i = 1)}$   is computed  in recursive steps that can be illustrated on the building block. 
In Fig. \ref{fig:Tannergraph2.1},  
\begin{equation}\label{eq:LLR_computation_upper}
    L_{u_{0}}(\mathbf{y})= L_{x_{0}}(y_{0}) \boxplus L_{x_{1}}(y_{1}),
\end{equation}
for  $i=0$,
where $L_{x_{0}}(y_{0})$ and $L_{x_{1}}(y_{1})$ are channel level LLRs and the box-plus operation between two LLR values $L_0$ and $L_1$ is defined as $L_0 \boxplus L_1 = \log \frac{1+ e^{L_0} e^{L_1}}{e^{L_0}+e^{L_1}}$. For $i=1$,
\begin{equation}\label{eq:LLR_computation_lower}
L_{u_{1}}(\mathbf{y},\hat{u}_0)=(-1)^{\hat{u}_{0}} L_{x_{0}}(y_{0}) + L_{x_{1}}(y_{1}),					
\end{equation} 
with the bit value $\hat{u}_{0}$ available from the previous decoding stage.
For example, the LLR $L_{u_2}(\mathbf{y}_2)$ in Fig. \ref{fig:Tannergraph4} is computed using \eqref{eq:LLR_computation_upper} from the intermediate LLRs $L_{v_{1,1}}(.)$ and $L_{v_{3,1}}(.)$. 
The LLR $L_{v_{1,1}}(.)$ is in turn computed according to \eqref{eq:LLR_computation_lower} from the channel level LLRs $L_{x_0}(y_0)$ and $L_{x_1}(y_1)$ as well as the bit estimate $\hat{v}_{0,1}$.
The LLR $L_{v_{3,1}}(.)$ is computed in a similar fashion from $L_{x_2}(y_2)$,  $L_{x_3}(y_3)$ and $\hat{v}_{2,1}$.%

\subsection{Successive Cancellation List Decoding}\label{subsec:SCL}
The SCL decoder \cite{tal_list_2015} can be seen as multiple SC decoders working in parallel. Every time an estimate $\hat{u}_i$ for $i {\in} \mathcal{A}$ has to be made, the decoder proceeds as an SC decoder for both possible decisions of $\hat{u}_i$ instead of using \eqref{eq:SC_decoding_rule}. The number of decoding paths doubles at each decoding stage $i{\in} \mathcal{A}$.
If the number of decoding paths in the list exceeds $N_L$ at any stage, the decoder retains only the $N_L$ most likely decoding paths, dropping the rest.
The likelihood of the correctness of a path $l \in \left\lbrace 0,1\dots N_L-1 \right\rbrace $ in the list at stage $i$ is conveyed by the path metric $M_{i,l}$ \cite{balatsoukas-stimming_llr-based_2015}
\begin{equation}
	M_{i,l}=M_{i-1,l} +  \log (1+e^{ -(1-2\hat{u}_{i,l}) L_{u_i}(\mathbf{y}_{i,l}) } ),	
	\label{eq:path_metric}				
\end{equation}
where $M_{i-1,l}$ is the path metric of the $l$th path at decoding stage $i-1$, $\hat{u}_{i,l}$ is the bit value with which the path is being extended, and ${L_{u_i}(\mathbf{y}_{i,l}) }$ is the  LLR value  for the $l$th path  according to \eqref{eq:LLR_computation_upper} or \eqref{eq:LLR_computation_lower}.

After the last decoding stage  $i{=}N{-}1$, the most likely decoding path from the list, i.e., the one having the smallest path metric, is selected as the decoder output. 
In the CRC-aided settings, a CRC checksum of $N_\text{CRC}$ bits is appended to the $K$ information bits and the $K+N_\text{CRC}$ bits are  encoded into an $N$ bit codeword using \eqref{eq:NxN_generator_matrix}. The decoder output is then the most likely decoding path in the final list that passes the CRC check.
If no path passes the CRC test, the most likely path in the list is selected as the decoder output.

\section{Finite Alphabet Decoding}\label{sec:finitealphabetdecoders}

Finite alphabet decoders are a family of quantized decoders that replace LLRs with integer valued messages in order to achieve a reduced space complexity. Instead of exchanging exact or approximated LLRs, $w$-bit messages $t$ from a finite alphabet $\mathcal{T}$ of size $|\mathcal{T}|{=}2^w$ are used to convey the reliability information w.r.t. a certain bit $x$. Thus, each message $t$ corresponds to an LLR level $L_x(t)$.

A general choice for the finite alphabet $\mathcal{T}$ is unsigned integers $\{0,1,\dots, |\mathcal{T}|-1\}$, e.g., as in \cite{lewandowsky_information-optimum_2018,shah_design_2019}. 
 However, this work uses a symmetric finite alphabet $\mathcal{T}{=}\{{-}2^{w{-}1},\ldots, {-}1,{+}1 \ldots, {+}2^{w{-}1}\}$ that is convenient to describe the proposed simplified hardware implementation\cite{mohr_uniform_2022,mohr_variable_2022}. The alphabet  $\mathcal{T}$ is typically chosen such that it is  sorted w.r.t. the underlying LLRs, i.e., $L_x(t{=}{-}2^{w{-}1)}<\ldots<L_x(t{=}{+}2^{w{-}1)}$. In the design of the decoder, the LLRs $L_x(t)$ are enforced to exhibit odd symmetry as
\begin{align}
|L_{x}(t)| = |L_{x}(-t)| \,\forall\, t\in \mathcal{T}.
\end{align}
The first half of such an alphabet translates to negative LLRs while the second half translates to positive LLR values.

The LLR computations in finite alphabet decoders are replaced with compression operations with some input $y{\in} \mathcal{Y}$ and output $t{\in}\mathcal{T}$ where $|\mathcal{T}|{<}|\mathcal{Y}|$. 
In order to minimize the loss in error correction performance of the decoder under the constrained resolution $w$, a mutual information maximizing decoder design aims at $\max_{p(t|y)}I(X;T)$ when designing the operations.
This kind of situation is classified as an information bottleneck setup where $X$ is the relevant, $Y$ is the observed and $T$ is the compressed variable\cite{tishby2000information}.
The information bottleneck framework  provides algorithms for determining the mapping $p(t|y)$ as well as the output joint distribution $p(x,t)$ from an input joint distribution $p(x,y)$. 
The mapping $p(t|y)$ is designed by placing $|\mathcal{T}|{-}1$ boundaries in the sorted observed alphabet $\mathcal{Y}$ and optimizing them to maximize $I(X;T)$. 
The distribution $p(x,t)$ is used to obtain the LLRs $L_{x}(t)$ and the distribution $p(t)$ of the compressed messages.
The deterministic mapping $p(t|y)$ represents the compression operation in the form of a lookup table.

\subsection{Mutual Information Maximizing Polar Decoders}

In finite alphabet polar decoders the information bottleneck method can be used to construct lookup tables which replace \eqref{eq:LLR_computation_upper} and \eqref{eq:LLR_computation_lower}\cite{shah_design_2019,shah_coarsely_2019}. This process is recapped here for the building block of Fig. \ref{fig:Tannergraph2.1} where the underlying channel $p(y_i|x_i)$ is a quantized binary input AWGN channel. 

Construction of the decoding lookup table  begins by designing a mutual information maximizing channel quantizer such that $y_i{\in} \mathcal{T}$\cite{lewandowsky_information-optimum_2018}. 
With the quantized channel outputs $\mathbf{y}{=}[y_0, y_1]^T$  at hand, the lookup table $p(t_0|\mathbf{y})$ is designed for the upper branch update with $t_0{\in}\mathcal{T}$ which compresses the input alphabet of size $2^{2w}$ to an output alphabet of size $2^w$.  Similarly, the lookup table $p(t_1|\mathbf{y},\hat{u}_0)$ is designed for the lower branch update with $t_1{\in}\mathcal{T}$ which compresses the input alphabet of size $2^{2w+1}$ to an output alphabet of size $2^w$. Both  $t_0$ and $t_1$ can be translated to LLR values $L_{u_0}(t_0)$ and $L_{u_1}(t_1)$, respectively. 
The mappings $p(t_0|\mathbf{y})$ and $p(t_1|\mathbf{y},\hat{u}_0)$ define a non uniform quantization of the underlying LLR space of thier inputs.

For a polar code of length $N$, there are $N-1$ distinct decoding tables for upper branch updates as well as $N-1$ distinct tables for the lower branch updates \cite{shah_design_2019,shah_coarsely_2019}. For instance, the decoder for Fig. \ref{fig:Tannergraph4} requires $2N{-}2{=}6$ distinct decoding tables: A common decoding table for both the upper branch updates at level $j{=}0$ and an individual decoding table for each upper branch update at the level $j{=}1$. Similarly,  a single decoding table for both the lower branch updates at level $j{=}0$ and a decoding table for each lower branch update at level $j{=}1$. Each upper branch decoding table has a size of $2^{2w}$ while each lower branch decoding table is of size $2^{2w+1}$. For further details, the reader is referred to \cite{shah_design_2019,shah_coarsely_2019,shah_MSIB_2023}.

\section{Proposed Efficient Decoder Implementation}\label{sec:proposeddecoder}

A key challenge in finite alphabet decoders is the efficient implementation of the mutual information maximizing lookup tables. In that regard, the computational domain implementation of the lookup tables in \cite{he_mutual_2019} offers an elegant solution for LDPC decoders which is adopted for polar decoders here. %

Recall that (\ref{eq:LLR_computation_upper}) and (\ref{eq:LLR_computation_lower}) deliver the result of the upper and lower branch update as $L_{u_0}(\mathbf{y})$ and $L_{u_1}(\mathbf{y},\hat{u}_0)$, respectively. For avoiding expensive propagation of the high resolution message to the building blocks of the next level in the code structure, quantization of the two LLRs is indispensable.
Consider an observed variable $Y,y{\in}\mathcal{Y}$ that models a high resolution LLR related to a relevant binary variable $X, x{\in}\mathcal{X}$. It can be shown that threshold quantization of $Y$ to a compressed variable $T,t{\in}\mathcal{T}$ using a set of thresholds $\tau$ can maximize the preserved mutual information $\max_{\mathbf{\tau}}I(X;T)$\cite{kurkoski_quantization_2014}. 
While the decoders designed in \cite{shah_design_2019,lewandowsky_information-optimum_2018} with the information bottleneck method use the result of such a threshold quantization in the form of a lookup table, \cite{he_mutual_2019} uses these thresholds for performing the quantization in a computational domain. In other words, the boundaries or thresholds determined during the lookup table design are used for implementing compression operations. Such a threshold quantization is henceforth represented as $t=Q(y)$.

In order to simplify the implementation, symmetric quantization is considered where the sign is preserved and the magnitude is clustered using thresholds $\mathcal{\tau}=\{\tau_0,\ldots,\tau_{2^{w-1}-2}\}$ in the following non-uniform quantization\cite{mohr_uniform_2022}:
\begin{align}
Q(y){=}\operatorname{sgn}(y)\begin{dcases}
1 & |y|{\leq}\tau_{0}\\
i & \tau_{i-2}{<}|y|{\leq}\tau_{i-1}, 2{<}i{<}2^{w-1}{-}1\\
2^{w-1} & |y|{>}\tau_{2^{w-1}-2} 
\end{dcases}
\label{equ:nonuniform_quanitzation}
\end{align}
For building block of polar codes in Fig. \ref{fig:Tannergraph2.1}, we have $\mathcal{X}{=}\mathcal{U}_0$ and $\mathcal{Y}{=}\{L_{u_0}(\mathbf{y}):\mathbf{y}{\in}\mathcal{Y}_0{\times}\mathcal{Y}_1\}$ for the upper branch. For the lower branch we have $\mathcal{X}{=}\mathcal{U}_1$ and $\mathcal{Y}{=}\{L_{u_1}(\mathbf{y},u_0):\mathbf{y}{\in}\mathcal{Y}_0{\times}\mathcal{Y}_1\And u_0{\in}\mathcal{U}_0\}$.

\subsection{Upper Branch Update}
The mutual information maximizing update for the upper branch leads to $t_0{=}Q(L_{u_0}(\mathbf{y}))$. It can be implemented as a lookup table like in \cite{shah_coarsely_2019, shah_design_2019, shah_MSIB_2023} or alternatively as a computation with quantization (only done for LDPC codes yet)\cite{he_mutual_2019,mohr_coarsely_2021,mohr_uniform_2022}.
Another, very hardware friendly solution is to approximate (\ref{eq:LLR_computation_upper}) using the so-called min-sum rule. By making use of the symmetric alphabet $\mathcal{T}$, no translation to LLRs is required and it naturally preserves the desired $w$-bit message resolution: 
\begin{align}
    t_0=\operatorname{sgn}(y_0)\operatorname{sgn}(y_1)\min(|y_0|, |y_1|) 
    \label{eq:minsum_update}
\end{align}
The approximation causes only minor performance degradation as shown in \cite{shah_MSIB_2023} and is therefore the recommended choice for the upper branch update.
\subsection{Lower Branch Update}
The mutual information maximizing update for the lower branch leads to $t_1{=}Q(L_{u_1}(\mathbf{y},\hat{u}_0))$. It can be implemented as a lookup table like in \cite{shah_coarsely_2019, shah_design_2019, shah_MSIB_2023} or alternatively as a computation with threshold quantization (only done for LDPC decoders yet) as in \cite{he_mutual_2019,mohr_coarsely_2021,mohr_uniform_2022}. The lookup table implementation suffers from its large size to cover all the $2^{2w+1}$ input combinations. This aspect is significantly improved when using the computation according to (\ref{eq:LLR_computation_lower}).

Up to this point, the operation's internal computations have been considered with real valued numbers. For a hardware implementation this is not acceptable. To reduce the internal resolution one option is to scale the real valued LLRs to an integer range from $-\iota$ to $+\iota$, with $\iota=2^{w'-1}{-}1$, as follows:
\begin{align}
    \phi_{s}(t)=\left\lceil s L(t)\right\rfloor=\operatorname{sgn}(L(t))\min\left(\left\lfloor s|L(t)|+0.5\right\rfloor,\iota\right)
\end{align}
where the scaling $s\in \mathbb{R}$ controls the LLR resolution $\Delta=1/s$ in the integer domain. Then, the integer computation yields
\begin{align}
        t_1=Q\left((-1)^{\hat{u}_{0}} \phi_{s}(y_{0}) + \phi_{s}(y_{1})\right)\approxeq Q(L_{u_1}(\mathbf{y},\hat{u}_0)).
\end{align}
\begin{table}[t]
    \centering
    \setlength{\tabcolsep}{1.6pt}
    \vspace{0.3cm}
    \caption{Accurate binary conversions for $w^\prime=3$.}
    \vspace{-0.15cm}
    \begin{tabular}{c|cc|cc|cc|cc|cc|cc|cc|cc}  
     $a$&+0&000&+1&001&+2&010&+3&011&-3&111&-2&110&-1&101&-0&100\\
    $\vartheta_{\text{2's}}(a)$&+0&000&+1&001&+2&010&+3&011&-3&101&-2&110&-1&111&+0&000\\
    \hline
    $b$&+0&000&+1&001&+2&010&+3&011&-4&100&-3&101&-2&110&-1&111\\
    $\vartheta_{\text{SM}}(b)$&+0&000&+1&001&+2&010&+3&011&-&-&-3&111&-2&110&-1&101
    \end{tabular}
    \label{tab:binary_format_conversion}
    \vspace{0.3cm}
    \centering
    \caption{Simplified binary conversions $w'=3$.}
    \vspace{-0.15cm}
    \setlength{\tabcolsep}{1.6pt}
    \begin{tabular}{c|cc|cc|cc|cc|cc|cc|cc|cc}  
     $a$&+0&000&+1&001&+2&010&+3&011&-3&111&-2&110&-1&101&-0&100\\
    $\vartheta_{\text{2's}}(a)$&+0&000&+1&001&+2&010&+3&011&-4&100&-3&101&-2&110&-1&111\\
    \hline
    $b$&+0&000&+1&001&+2&010&+3&011&-4&100&-3&101&-2&110&-1&111\\
    $\vartheta_{\text{SM}}(b)$&+0&000&+1&001&+2&010&+3&011&-&-&-2&110&-1&101&-0&100
    \end{tabular}
    \vspace{-0.4cm}
    \label{tab:binary_format_conversion_approximated}
\end{table}
\def\unit{0.78cm}
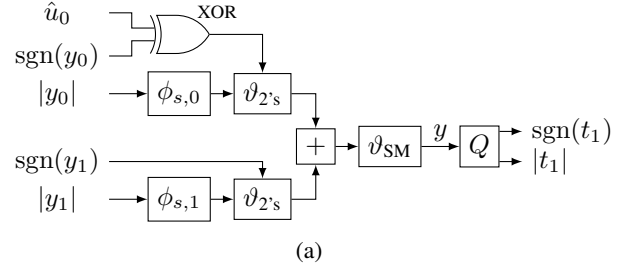
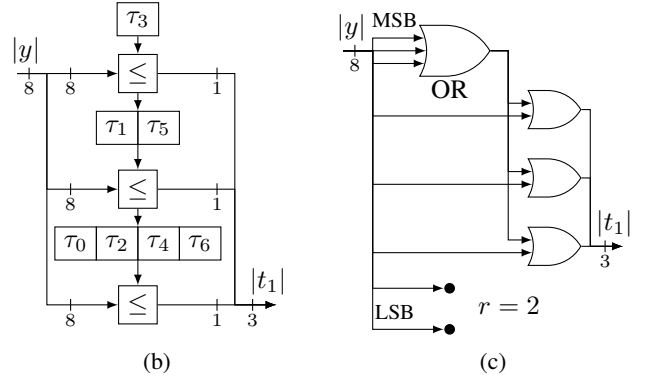
\begin{figure}[t]
\centering  
        \begin{subfigure}[b]{0.9\linewidth}
        	\begin{tikzpicture}[yscale=0.4, xscale=1, node distance=0.3cm, auto]
	
		\def \nodesize {0.5} %
		\def \VertDist {2.5}
		\def \HorDist {1}
		\def \HorDisplace {0.5}
            \def \myunit {1.0cm}

            \node(u0hat) at (0,0)[]{$\hat{u}_0$};
            \node(sgn_y0)[below=of u0hat, yshift=0.3*\myunit]{$\operatorname{sgn}(y_0)$};
            \node(mag_y0)[below=of sgn_y0, yshift=0.4*\myunit]{$|y_0|$};

            \node(sgn_y1)[below=of mag_y0]{$\operatorname{sgn}(y_1)$};
            \node(mag_y1)[below=of sgn_y1, yshift=0.4*\myunit]{$|y_1|$};

            \node(lut0)[draw, rectangle, xshift=0.5*\myunit,right=of mag_y0]{$\phi_{s,0}$};
            \coordinate(helperpoint) at ($(u0hat)!0.5!(sgn_y0)$);
            \node(xor0) at (helperpoint-|lut0)[xor gate US, draw, logic gate inputs=nn]{};
            \node(lut1) at (mag_y1-|lut0)[draw, rectangle]{$\phi_{s,1}$};

            \node(twos0)[draw, rectangle, right=of lut0]{$\vartheta_\text{2's}$};
            \node(twos1)[draw, rectangle, right=of lut1]{$\vartheta_\text{2's}$};

            \node(sum) at ($(twos0)!0.5!(twos1)$)[draw, rectangle, xshift=0.7*\myunit]{$+$};

            \node(sm) [draw, rectangle, right=of sum]{$
            \vartheta_\text{SM}$};

            \node(quant)[draw, rectangle, right=of sm, xshift=0.2*\myunit]{$Q$};
            
            \node(out_sgn)[right=of quant, yshift=0.2*\myunit]{$\operatorname{sgn}(t_1)$};
            \node(out_mag)[right=of quant, yshift=-0.2*\myunit]{$|t_1|$};

            \draw(sgn_y0.east)-|($(sgn_y0.east)!0.5!(xor0.input 2)$)|-(xor0.input 2);
            \coordinate(sgn_y0_anchor) at (u0hat.east-|sgn_y0.east);
            \draw(sgn_y0_anchor)-|($(sgn_y0_anchor)!0.5!(xor0.input 1)$)|-(xor0.input 1);
            
            \draw[-latex](mag_y0-|sgn_y0.east)--(lut0.west);
            \draw[-latex](mag_y1-|sgn_y1.east)--(lut1.west);

            \draw[-latex](xor0.output)-|(twos0.north);
            \draw[-latex](lut0.east)--(twos0.west);

            \draw[-latex](sgn_y1.east)-|(twos1.north);
            \draw[-latex](lut1.east)--(twos1.west);

            \draw[-latex](twos0)-|(sum.north);
            \draw[-latex](twos1)-|(sum.south);
            \draw[-latex](sum.east)--(sm.west);
            \draw[-latex](sm)--node[above, rotate=0, xshift=0.0*\myunit,yshift=-0.05*\myunit,scale=1.0]{$y$}(quant.west);%
            \draw[-latex](quant.east|-out_sgn.west)--(out_sgn.west);
            \draw[-latex](quant.east|-out_mag.west)--(out_mag.west);
            
		\node(labelxor) at (xor0)[yshift=0.3*\myunit,xshift=0.5*\myunit, scale=0.7]{XOR};	
	\end{tikzpicture}
        \vspace{-0.4cm}
        \caption{}
    \vspace{0.1cm}
        \label{fig:hardware_lower_branch}
    \end{subfigure}
    \begin{subfigure}[b]{0.47\linewidth}
         \begin{tikzpicture} [scale=1.0, x=\unit,y=\unit, node distance=0.3*\unit and 0.5*\unit]     
    \node(b3)[rectangle, draw=black]{$\tau_3$};
    \node(comp1)[rectangle, draw=black, below=of b3]{$\le$};
    
    \node(b1)[rectangle, draw=black, below=of comp1, xshift=-0.35*\unit]{$\tau_1$};
    \node(b5)[rectangle, draw=black, below=of comp1, xshift=0.35*\unit]{$\tau_5$};
    \node(comp2)[rectangle, draw=black, below=of comp1, yshift=-1*\unit]{$\le$};
    
    \node(b0)[rectangle, draw=black, below=of comp2, xshift=-1.05*\unit]{$\tau_0$};
    \node(b2)[rectangle, draw=black, below=of comp2, xshift=-0.35*\unit]{$\tau_2$};
    \node(b4)[rectangle, draw=black, below=of comp2, xshift=0.35*\unit]{$\tau_4$};
    \node(b6)[rectangle, draw=black, below=of comp2, xshift=1.05*\unit]{$\tau_6$};
    
    \node(comp3)[rectangle, draw=black, below=of comp2, yshift=-1*\unit]{$\le$};
    
    \node(y)[left=of comp1, xshift=-1.2*\unit, label={[xshift=0.3*\unit, yshift=-0.15*\unit]{$|y|$}}]{};
    \node(t)[right=of comp3, xshift=1.5*\unit, label={[xshift=-0.3*\unit, yshift=-0.15*\unit]{$|t_1|$}}]{};
    
    \draw[-latex](b3.south)--(comp1.north);
    \draw[-latex](b1.south-|comp2.north)--(comp2.north);
    \draw[-latex](b0.south-|comp3.north)--(comp3.north);
    
    \draw[-latex](comp1.south)--(comp1.south|-b1.north);
    \draw[-latex](comp2.south)--(comp2.south|-b0.north);
    
    \draw[-latex](y.east)--(comp1.west);
    \draw[-latex](y.east)--+(0.5,0)|-(comp2.west);
    \draw[-latex](y.east)--+(0.5,0)|-(comp3.west);
    
    \draw[-latex](comp1.east)--+(1.3,0)|-(t.west);
    \draw[-latex](comp2.east)--+(1.3,0)|-(t.west);
    \draw[-latex](comp3.east)--(t.west);

    \coordinate (tbitwidth) at ($(t.west)+(-0.4*\unit, 0)$);
    \draw ($(tbitwidth)+(0, 0.1*\unit)$)--($(tbitwidth)-(0, 0.1*\unit)$)node[yshift=-0.15*\unit]{\scriptsize$3$};
    
    \coordinate (ybitwidth0) at ($(y.east)+(0.2*\unit, 0)$);
    \draw ($(ybitwidth0)+(0, 0.1*\unit)$)--($(ybitwidth0)-(0, 0.1*\unit)$)node[yshift=-0.15*\unit]{\scriptsize$8$};
    
    \coordinate (ybitwidth1) at ($(comp1.west)+(-0.8*\unit, 0)$);
    \draw ($(ybitwidth1)+(0, 0.1*\unit)$)--($(ybitwidth1)-(0, 0.1*\unit)$)node[yshift=-0.15*\unit]{\scriptsize$8$};
    
    \coordinate (ybitwidth2) at ($(comp2.west)+(-0.8*\unit, 0)$);
    \draw ($(ybitwidth2)+(0, 0.1*\unit)$)--($(ybitwidth2)-(0, 0.1*\unit)$)node[yshift=-0.15*\unit]{\scriptsize$8$};
    
    \coordinate (ybitwidth3) at ($(comp3.west)+(-0.8*\unit, 0)$);
    \draw ($(ybitwidth3)+(0, 0.1*\unit)$)--($(ybitwidth3)-(0, 0.1*\unit)$)node[yshift=-0.15*\unit]{\scriptsize$8$};
    
    \coordinate (comp1outbitwidth) at ($(comp1.east)+(1*\unit, 0)$);
    \draw ($(comp1outbitwidth)+(0, 0.1*\unit)$)--($(comp1outbitwidth)-(0, 0.1*\unit)$)node[yshift=-0.15*\unit]{\scriptsize$1$};
    
    \coordinate (comp2outbitwidth) at ($(comp2.east)+(1*\unit, 0)$);
    \draw ($(comp2outbitwidth)+(0, 0.1*\unit)$)--($(comp2outbitwidth)-(0, 0.1*\unit)$)node[yshift=-0.15*\unit]{\scriptsize$1$};
    
    \coordinate (comp3outbitwidth) at ($(comp3.east)+(1*\unit, 0)$);
    \draw ($(comp3outbitwidth)+(0, 0.1*\unit)$)--($(comp3outbitwidth)-(0, 0.1*\unit)$)node[yshift=-0.15*\unit]{\scriptsize$1$};
\end{tikzpicture}	
        \caption{}
        \label{fig:hardware_quant_nonuniform}
    \end{subfigure}
    \begin{subfigure}[b]{0.5\linewidth}
         \begin{tikzpicture} [scale=1.0, x=\unit,y=\unit, node distance=0.5*\unit and 0.5*\unit]   
     \node(y)[yshift=1*\unit, label={[xshift=0.3*\unit, yshift=-0.15*\unit]{$|y|$}}]{};
     \node(orclip)[or gate US, draw, logic gate inputs=nnn, right=of y, xshift=0.9*\unit, label={[xshift=-0.0*\unit, yshift=-1.4*\unit]{OR}}]{};
     
     \node(orclip1)[or gate US, draw, logic gate inputs=nn, right=of y, xshift=2.7*\unit, yshift=-1*\unit]{};
     \node(orclip2)[or gate US, draw, logic gate inputs=nn, below=of orclip1]{};
     \node(orclip3)[or gate US, draw, logic gate inputs=nn, below=of orclip2]{};
     
     \node(deadend1)[circle,fill=black,inner sep=0pt,minimum size=4pt, below=of orclip, yshift=-3*\unit]{};
     \node(deadend2)[circle,fill=black,inner sep=0pt,minimum size=4pt, below=of deadend1]{};
     
     \node(t)[right=of orclip3, xshift=0.2*\unit, label={[xshift=-0.3*\unit, yshift=-0.15*\unit]{$|t_1|$}}]{};
     
     \draw[-latex](y.east)-|+(0.5,0)|-(orclip.input 1);
     \draw[-latex](y.east)-|+(0.5,0)|-(orclip.input 2);
     \draw[-latex](y.east)-|+(0.5,0)|-(orclip.input 3);
     
     \draw[-latex](orclip.east)-|+(0.3,0)|-(orclip1.input 1);
     \draw[-latex](y.east)-|+(0.5,0)|-(orclip1.input 2);
     
     \draw[-latex](orclip.east)-|+(0.3,0)|-(orclip2.input 1);
     \draw[-latex](y.east)-|+(0.5,0)|-(orclip2.input 2);
     
     \draw[-latex](orclip.east)-|+(0.3,0)|-(orclip3.input 1);
     \draw[-latex](y.east)-|+(0.5,0)|-(orclip3.input 2);
     
     \draw[-latex](y.east)-|+(0.5,0)|-(deadend1);
     \draw[-latex](y.east)-|+(0.5,0)|-(deadend2);
     
     \draw[-latex](orclip1.east)-|+(0.15,0)|-(t.west);
     \draw[-latex](orclip2.east)-|+(0.15,0)|-(t.west);
     \draw[-latex](orclip3.east)-|+(0.15,0)|-(t.west);
     
     \coordinate (ybitwidth0) at ($(y.east)+(0.2*\unit, 0)$);
     \draw ($(ybitwidth0)+(0, 0.1*\unit)$)--($(ybitwidth0)-(0, 0.1*\unit)$)node[yshift=-0.15*\unit]{\scriptsize$8$};
     
     \coordinate (tbitwidth) at ($(t.west)-(0.3*\unit, 0)$);
     \draw ($(tbitwidth)+(0, 0.1*\unit)$)--($(tbitwidth)-(0, 0.1*\unit)$)node[yshift=-0.15*\unit]{\scriptsize$3$};
     
     \node (MSB) at ($(orclip.input 1)+(-0.5*\unit, 0.3*\unit)$){\footnotesize MSB};
     \node (LSB) at ($(deadend2.west-|MSB)+(0, 0.3*\unit)$){\footnotesize LSB};
     
     \node (rshift) at ($(deadend1)+(1*\unit, -0.3*\unit)$){$r=2$};
\end{tikzpicture}
        \caption{}
        \label{fig:hardware_quant_uniform}
    \end{subfigure}
    \caption{(a) shows a hardware schematic for the lower branch processing that can be used with (b) non-uniform quantization or (c) uniform quantization\cite{mohr_uniform_2022}. For (b) and (c) we have $w'{=}9$-bit internal resolution and $w{=}4$-bit message resolution.}
    \label{fig:hardware_quantization}
   \vskip -10pt
\end{figure}
Fig.~\ref{fig:hardware_lower_branch} depicts a corresponding hardware schematic where the translations are assumed to be implemented with two $(w{-}1)$-bit lookup tables. The adder is assumed to work in a binary two's complement format such that subtraction and addition can be performed with the same hardware module. The quantization expects a sign-magnitude format. Therefore, two conversions from sign magnitude into the 2's complement format $b{=}\vartheta_\text{2's}(a)$ and vice versa $a{=}\vartheta_\text{SM}(b)$ must be part of the hardware. Table \ref{tab:binary_format_conversion} describes an accurate conversion
\begin{subequations}\label{equ:accurate_binary_conversion}
    \begin{align}
    b=\vartheta_{\text{2's}}(a)&=[a_0,(|a|\oplus a_0) +(+1_{\text{2's}}\land a_0)] \text{ and }\\
    a=\vartheta_{\text{SM}}(b)&=[b_0,(|b|+(-1_{\text{2's}}\land b_0))\oplus b_0]
\end{align}
\end{subequations}
where $a_0$ ($b_0$) refers to sign bit of a number $a$ ($b$), $+$ is binary addition with carry propagation and, eventually, $\land$ and $\oplus$ are bitwise logic AND and XOR operations.
In particular the $+$ operation causes significant complexity. Therefore, an approximated conversion is proposed according to 
\begin{subequations}\label{equ:simple_binary_conversion}
    \begin{align}
    b=\vartheta_{\text{2's}}(a)&=[a_0, |a|\oplus a_0] \text{ and }\\ 
    a=\vartheta_{\text{SM}}(b)&=[b_0, |b|\oplus b_0]    
\end{align}
\end{subequations}
which is illustrated in Table~\ref{tab:binary_format_conversion_approximated}.
The technique involves a slight bias, since e.g. $\vartheta({+}1_{2's}){=}{{+}1}_{\mathrm{SM}}$ but $\vartheta({{-}1}_{2's}){=}{{-}2}_{\mathrm{SM}}$. To distribute the bias fairly, one option is to let every second lower branch update invert the sign for inputs and output. 
In our simulations we used the accurate variant (\ref{equ:accurate_binary_conversion}) but we expect only insignificant performance loss from the much simpler conversion (\ref{equ:simple_binary_conversion}).

Another bottleneck is the computation of (\ref{equ:nonuniform_quanitzation}). It requires $w{-}1$ threshold comparisons when being implemented in a binary search manner, as depicted in Fig.~\ref{fig:hardware_quant_nonuniform}.
As proposed in \cite{mohr_uniform_2022} for LDPC codes, a restriction to uniform thresholds enables a much simpler implementation of the quantization operation, which is shown in Fig.~\ref{fig:hardware_quant_uniform}. In that approach the quantization is achieved by a clipping and bit shifting operation defined as
\begin{align}
    Q(y) = \operatorname{sgn}(y) \min(\lfloor|y|/2^r\rfloor + 1, 2^{w-1})
\end{align}
where $r$ denotes the number of right-shifted bit positions. 
By modifying $r$ and the scaling factor $s=1/\Delta$ for the translation tables, any uniform threshold spacing, $(\tau_{i+1}-\tau_{i})=\Delta2^{r}$, can be achieved. 
The optimal uniform quantization is obtained with a grid based search aiming for $\max_{s,r} I(U_1; T_1)$.
\subsection{Complexity Analysis}
\begin{table}[t]
    \centering
    \vspace{0.1cm}
    \caption{Complexity of lower branch updates.}%
    \label{table:complexity_lower_branch}
    \setlength{\tabcolsep}{3pt}
    \begin{tabular}{c|c|cccc}
        \multirow{2}{*}{\shortstack[c]{variant}}& \multirow{2}{*}{\shortstack[c]{additions/\\comparisons}}  & \multicolumn{3}{c}{\shortstack[c]{memory usage in bit (e.g. $w'{=}6$)}} \\
         &&general&$w{=}4$&$w{=}3$&$w{=}2$\\
        \hline
        IB-LUT &0  & $w\cdot2^{2w+1}$&2048&384&64\\
        CD (non-uni.) &$w$  &$(2(w'{-}1)+w')2^{w-1}$&128&64&32\\
        CD (uniform) & $1$    &$2(w'{-}1)2^{w-1}$ &80 &40 & 20
    \end{tabular}
    \vspace{-0.3cm}
\end{table}
For the upper branch processing the lowest complexity is observed with the min-sum update which requires only a single exclusive-or gate and a $(w{-}1)$-bit comparison (see (\ref{eq:minsum_update})).

A comparison of the complexity for the discussed lower branch updates is provided in Table~\ref{table:complexity_lower_branch}. The total number of potentially different parameterized updates is $N{-}1$ for the complete decoder. The example memory usage is evaluated for an internal resolution of $w'{=}6$\,bits which sacrifices only minor performance in the simulations. 

Clearly, the computational domain solution with uniform quantization yields the lowest complexity. It only requires memory for the two translations to $w'$-bit LLR magnitudes of and one addition operation. The non-uniform computational domain variant requires additional complexity to perform the non-uniform threshold quantization with $w{-}1$ comparisons tested against the total $2^{w-1}$ different $w'$-bit thresholds in a binary search fashion.
In case of $w{=}4$\,bits the lookup table solution requires more than $25$ times the number of memory bits to specify the decoder.
A conventional fixed point SC decoder from \cite{balatsoukas-stimming_llr-based_2015} calls for about $w{=}6$\,bits to achieve similar performance as the proposed $w{=}4$-bit decoder.

\section{Performance Analysis}\label{sec:performanceevaluation}
This section presents the simulation results showing the error correction performance of the proposed quantized decoders. 
The proposed decoding scheme is compared with double-precision floating-point LLR-based decoding as well as finite alphabet decoders designed using the information bottleneck method \cite{shah_design_2019,shah_coarsely_2019}. 
The LLR-based  decoding represents the unquantized decoders. The SCL decoding is used here in the CRC-aided setting with list size of $N_L=32$ and CRC size of $N_{\mathrm{CRC}}=16$.
For the construction of polar codes, the method adopted in 5G NR \cite{3gpp_38_212} is used.
Finally, all the simulations are performed for a codeword length of $N=1024$ over an AWGN channel using BPSK modulation.

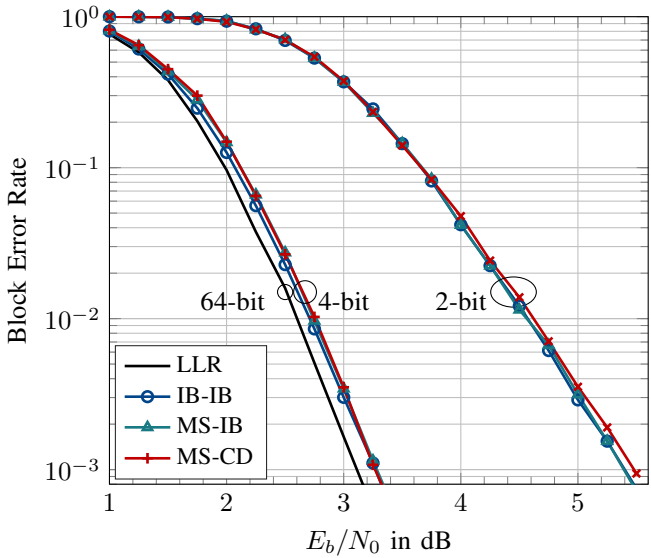
\begin{figure}[tb]
	\centering
	\begin{tikzpicture}
\pgfplotsset{legend style={font=\small}}
\begin{axis}[%
width=0.8*\columnwidth,
height=0.7*\columnwidth,
scale only axis,
xmin=1,
xmax=5.6,
ymode=log,
ymin=8e-04,
ymax=1,
yminorticks=true,
ylabel={Block Error Rate},
xlabel={$E_b/N_0$ in dB},
axis background/.style={fill=white},
xmajorgrids,
ymajorgrids,
yminorgrids,
xminorticks=true, minor x tick num=4,
legend style={at={(0.01,0.01)}, anchor=south west},
legend style={legend cell align=left, align=left, draw=white!15!black}
]
\addplot [color=black,solid,line width=1.0pt]
table{%
0	0.997000000000000
0.250000000000000	0.987666666666667
0.500000000000000	0.967500000000000
0.750000000000000	0.889000000000000
1	0.766500000000000
1.25000000000000	0.581000000000000
1.50000000000000	0.380000000000000
1.75000000000000	0.203500000000000
2	0.0964166666666667
2.25000000000000	0.0375714285714286
2.50000000000000	0.0159218750000000
2.75000000000000	0.00508629441624366
3	0.00165016501650165
3.25000000000000	0.000543478260869565
3.50000000000000	0.000170829170829171
3.75000000000000	6.59340659340659e-05
};
\addlegendentry{LLR};
\addplot [color=darkblue,solid,mark=o,line width=1.0pt]
table{%
0	0.999000000000000
0.250000000000000	0.988500000000000
0.500000000000000	0.974000000000000
0.750000000000000	0.905500000000000
1	0.803000000000000
1.25000000000000	0.610500000000000
1.50000000000000	0.416000000000000
1.75000000000000	0.247666666666667
2	0.126000000000000
2.25000000000000	0.0560000000000000
2.50000000000000	0.0227777777777778
2.75000000000000	0.00854700854700855
3	0.00300898203592814
3.25000000000000	0.00110485651214128
3.50000000000000	0.000352138871667700
};
\addlegendentry{IB-IB};%

\addplot [color=darkgreen,solid,mark=triangle,line width=1.0pt]
table{%
0	0.999000000000000
0.250000000000000	0.993500000000000
0.500000000000000	0.976500000000000
0.750000000000000	0.923500000000000
1	0.810000000000000
1.25000000000000	0.636500000000000
1.50000000000000	0.433500000000000
1.75000000000000	0.282000000000000
2	0.147200000000000
2.25000000000000	0.0662222222222222
2.50000000000000	0.0273809523809524
2.75000000000000	0.00967241379310345
3	0.00343558282208589
3.25000000000000	0.00114553014553015
3.50000000000000	0.000405007363770250
};
\addlegendentry{MS-IB}; %

\addplot [color=darkred,solid,mark=+,line width=1.0pt]
table{%
0	1
0.250000000000000	0.993000000000000
0.500000000000000	0.975000000000000
0.750000000000000	0.920000000000000
1	0.819000000000000
1.25000000000000	0.648000000000000
1.50000000000000	0.449000000000000
1.75000000000000	0.300000000000000
2	0.148666666666667
2.25000000000000	0.0650000000000000
2.50000000000000	0.0266666666666667
2.75000000000000	0.0102666666666667
3	0.00351162790697674
3.25000000000000	0.00107500000000000
3.50000000000000	0.000396301188903567
};
\addlegendentry{MS-CD}; %

\addplot [color=darkblue,solid,mark=o,line width=1.0pt]
table{%
0	1
0.250000000000000	1
0.500000000000000	1
0.750000000000000	1
1	1
1.25000000000000	0.997500000000000
1.50000000000000	0.995000000000000
1.75000000000000	0.970000000000000
2	0.933500000000000
2.25000000000000	0.831000000000000
2.50000000000000	0.696000000000000
2.75000000000000	0.531500000000000
3	0.370000000000000
3.25000000000000	0.245000000000000
3.50000000000000	0.144200000000000
3.75000000000000	0.0816250000000000
4	0.0418666666666667
4.25000000000000	0.0224285714285714
4.50000000000000	0.0121176470588235
4.75000000000000	0.00614285714285714
5	0.00289473684210526
5.25000000000000	0.00154241645244216
5.50000000000000	0.000746583850931677
};

\addplot [color=darkgreen,solid,mark=triangle,line width=1.0pt]
table{%
0	1
0.250000000000000	1
0.500000000000000	1
0.750000000000000	0.999500000000000
1	1
1.25000000000000	0.999500000000000
1.50000000000000	0.994000000000000
1.75000000000000	0.971000000000000
2	0.938500000000000
2.25000000000000	0.821000000000000
2.50000000000000	0.707500000000000
2.75000000000000	0.533000000000000
3	0.369500000000000
3.25000000000000	0.230000000000000
3.50000000000000	0.144833333333333
3.75000000000000	0.0841250000000000
4	0.0416875000000000
4.25000000000000	0.0224285714285714
4.50000000000000	0.0114500000000000
4.75000000000000	0.00667391304347826
5	0.00314583333333333
5.25000000000000	0.00154358974358974
5.50000000000000	0.000719424460431655
};

\addplot [color=darkred,solid,mark=x,line width=1.0pt]
table{%
0	1
0.250000000000000	1
0.500000000000000	1
0.750000000000000	1
1	1
1.25000000000000	0.997000000000000
1.50000000000000	0.992250000000000
1.75000000000000	0.968250000000000
2	0.924500000000000
2.25000000000000	0.823250000000000
2.50000000000000	0.702750000000000
2.75000000000000	0.541750000000000
3	0.374833333333333
3.25000000000000	0.234222222222222
3.50000000000000	0.140153846153846
3.75000000000000	0.0835238095238095
4	0.0477428571428571
4.25000000000000	0.0242461538461538
4.50000000000000	0.0138508771929825
4.75000000000000	0.00707623318385650
5	0.00353636363636364
5.25000000000000	0.00190256410256410
5.50000000000000	0.000942523136872869
};
\draw (axis cs:2.5,1.5e-2) ellipse (0.1cm and 0.1cm);
\node at (axis cs:2.05,1.3e-2){64-bit};

\draw (axis cs:2.67,1.5e-2) ellipse (0.15cm and 0.15cm);
\node at (axis cs:3.0,1.3e-2){4-bit};

\draw (axis cs:4.45,1.5e-2) ellipse (0.3cm and 0.2cm);
\node at (axis cs:4.0,1.3e-2){2-bit};
\end{axis}
\end{tikzpicture}%
	\caption{Block error rate comparison under SC decoding. 
 }
 \vspace{-0.5cm}
	\label{fig:BLER_SC}
\end{figure}

The SC decoding of polar codes mainly performs two types of computations, i.e., upper  or lower branch on a building block in the polar code structure. Hence, the finite alphabet decoders in this work are labelled according to the design method used for upper and lower branch updates. The decoders from \cite{shah_design_2019,shah_coarsely_2019} where both upper and lower branch updates are designed using the information bottleneck (IB) method are labelled IB-IB. The finite alphabet decoders of \cite{shah_MSIB_2023} that deploys min-sum (MS) and the information bottleneck for designing upper and lower branch updates, respectively, are labelled MS-IB. The proposed decoders which use min-sum rule for upper branch and computational-domain uniform quantization method for lower branch updates are labelled \mbox{MS-CD}. 

The finite alphabet quantized decoders are constructed offline for a selected $w {=} \log_2 ( |\mathcal{T}|)$-bit resolution. Each $w$-bit quantized decoder deploys a $w$-bit mutual information maximizing channel quantizer designed using the information bottleneck method \cite{shah_design_2019,shah_coarsely_2019,lewandowsky_information-optimum_2018}. The channel quantizer and, in turn, the quantized decoder are constructed for a specific $E_b/N_0$, which is referred to as the design $E_b/N_0$ of the decoder.
For a given code rate $R$ and resolution $w$, the design $E_b/N_0$ for the IB-IB decoder is selected as the one which achieves a block error rate of $10^{-3}$ at the smallest channel $E_b/N_0$. The same design $E_b/N_0$ is then used to generate MS-IB and MS-CD decoders for the same $R$ and $w$.

\subsection{Successive Cancellation Decoding}
Fig. \ref{fig:BLER_SC} shows the block error rates under the successive cancellation decoding for a code rate $R{=}0.5$ and resolution of $w{=}4$ and $2$ bits. The three finite alphabet decoders in the figure for $w{=}4$ bit resolution were designed for $E_b/N_0{=}0.5$\,dB. The $w{=}2$ bit decoders were design for $E_b/N_0{=}3.5$\,dB. Compared to the floating-point LLR-based decoder, the 4-bit decoder show a degradation of around 0.2\,dB while the 2-bit quantized decoders exhibit a significant performance loss of approximately 2.4 dB. Most importantly, the IB-IB, MS-IB and the proposed MS-CD decoders have practically the same error rate performance. Thus, the implementation friendly \mbox{MS-CD} approximation costs nothing in terms of performance loss.

\begin{figure}[tb]
	\centering
	\begin{tikzpicture}
\pgfplotsset{legend style={font=\small}}
\begin{axis}[%
width=0.8*\columnwidth,
height=0.7*\columnwidth,
scale only axis,
xmin=-0,
xmax=3.4,
ymode=log,
ymin=8e-04,
ymax=1,
yminorticks=true,
ylabel={Block Error Rate},
xlabel={$E_b/N_0$ in dB},
axis background/.style={fill=white},
xmajorgrids,
ymajorgrids,
yminorgrids,
xminorticks=true, minor x tick num=4,
legend style={at={(0.01,0.01)}, anchor=south west},
legend style={legend cell align=left, align=left, draw=white!15!black}
]
\addplot [color=black,solid,mark=.,line width=1.0pt]
table{%
0					0.96552
0.25					0.86318
0.5					0.70268
0.75					0.43804
1					0.20766
1.25					0.071776
1.5					1.76E-02
1.75					3.12E-03
2					3.78E-04
2.25					3.38E-05
2.5					2.75E-06
	
};
\addlegendentry{LLR};%

\addplot [color=darkblue,solid,mark=o,line width=1.0pt]
table[]{%
0	0.975114915200507
0.250000000000000	0.898232881035027
0.500000000000000	0.723400936037442
0.750000000000000	0.497574287446938
1	0.254123144584937
1.25000000000000	0.105874705452182
1.50000000000000	0.0317270934281331
1.75000000000000	0.00636124862083443
2	0.00117467286307353
2.25000000000000	0.000183000000000000
};
\addlegendentry{IB-IB};%

\addplot [color=darkgreen,solid,mark=triangle,line width=1.0pt]
table[]{%
0	0.978306508047586
0.250000000000000	0.911326602019394
0.500000000000000	0.762271318604419
0.750000000000000	0.527841647505748
1	0.294411676497051
1.25000000000000	0.122492252487359
1.50000000000000	0.0379098658753019
1.75000000000000	0.00816119455335780
2	0.00141571190317096
2.25000000000000	0.000199333333333333
};
\addlegendentry{MS-IB};%

\addplot [color=darkred,solid,mark=x,line width=1.0pt]
table{%
0	0.974675108297234
0.250000000000000	0.912362545818061
0.500000000000000	0.773408863712096
0.750000000000000	0.540653115628124
1	0.297734088637121
1.25000000000000	0.129123625458181
1.50000000000000	0.0414631091068521
1.75000000000000	0.00991265089802729
2	0.00165780316459852
2.25000000000000	0.000292508594250034
};
\addlegendentry{MS-CD};%

\addplot [color=black,solid,mark=.,line width=1.0pt]
table{%
0	0.364545151616128
0.250000000000000	0.203432189270243
0.500000000000000	0.0939597315436242
0.750000000000000	0.0339305711086226
1	0.0104450420751280
1.25000000000000	0.00247430137977238
1.50000000000000	0.000466990239740706
};

\addplot [color=darkblue,solid,mark=o,line width=1.0pt]
table{%
0	0.477840719760080
0.250000000000000	0.276741086304565
0.500000000000000	0.137120959680107
0.750000000000000	0.0596828567861156
1	0.0223600262357641
1.25000000000000	0.00670859538784067
1.50000000000000	0.00145381854527282
1.75000000000000	0.000376082037168017
};

\addplot [color=darkred,solid,mark=x,line width=1.0pt]
table{%
0	0.500499833388870
0.250000000000000	0.322392535821393
0.500000000000000	0.167610796401200
0.750000000000000	0.0779526400941315
1	0.0279983284580025
1.25000000000000	0.00866397982715145
1.50000000000000	0.00227407721680102
1.75000000000000	0.000494101509590732
};

\addplot [color=black,solid,mark=.,line width=1.0pt]
table{%
0	1
0.250000000000000	1
0.500000000000000	1
0.750000000000000	1
1	0.998500499833389
1.25000000000000	0.989170276574475
1.50000000000000	0.942019326891036
1.75000000000000	0.795401532822393
2	0.534155281572809
2.25000000000000	0.268910363212263
2.50000000000000	0.0873042319226924
2.75000000000000	0.0189488569408437
3	0.00287403573241440
3.25000000000000	0.000314870585839440
};

\addplot [color=darkblue,solid,mark=o,line width=1.0pt]
table{%
0	1
0.250000000000000	1
0.500000000000000	1
0.750000000000000	0.999833388870377
1	0.999166944351883
1.25000000000000	0.993835388203932
1.50000000000000	0.948850383205598
1.75000000000000	0.826724425191603
2	0.578307230923026
2.25000000000000	0.303232255914695
2.50000000000000	0.114961679440187
2.75000000000000	0.0301492537313433
3	0.00588526643941380
3.25000000000000	0.000856382166901253
};

\addplot [color=darkred,solid,mark=x,line width=1.0pt]
table{%
0	1
0.250000000000000	1
0.500000000000000	1
0.750000000000000	0.999888925913584
1	0.999111407308675
1.25000000000000	0.993113406642230
1.50000000000000	0.959902254803954
1.75000000000000	0.845051649450183
2	0.622237032100411
2.25000000000000	0.341552815728091
2.50000000000000	0.128957014328557
2.75000000000000	0.0400967531863429
3	0.00697026826020029
3.25000000000000	0.00109436821281253
3.50000000000000	0.000170372456907127
};

\draw (axis cs:0.75,5e-2) ellipse (0.3cm and 0.2cm);
\node at (axis cs:0.35,5e-2){$R{=}\frac{1}{4}$};

\draw (axis cs:1.4,5e-2) ellipse (0.3cm and 0.2cm);
\node at (axis cs:1.8,5e-2){$R{=}\frac{1}{2}$};

\draw (axis cs:2.6,5e-2) ellipse (0.3cm and 0.2cm);
\node at (axis cs:3.0,5e-2){$R{=}\frac{3}{4}$};
\end{axis}
\end{tikzpicture}%
	\caption{Block error rate comparison under CRC-aided SCL decoding with various code rates for $w=4$}
	\label{fig:BLER_4bit_multiple_rates}
 \vspace{-0.5cm}
\end{figure}
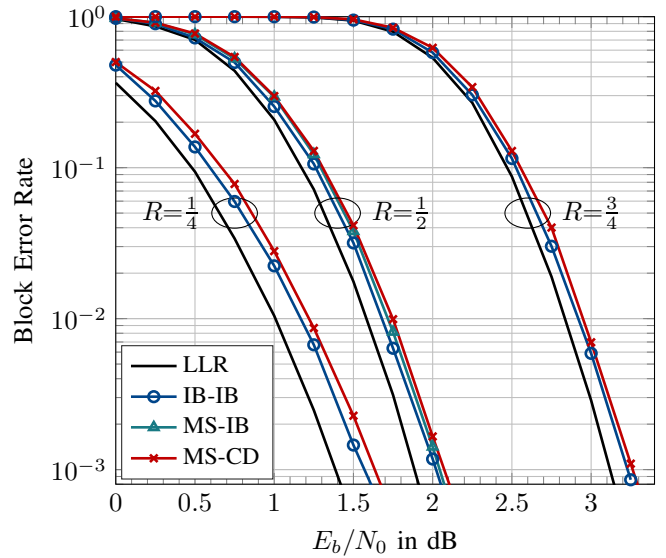

\subsection{Successive Cancellation List Decoding}
Fig. \ref{fig:BLER_4bit_multiple_rates} presents the block error rates for CRC-aided SCL decoding of 4-bit quantized decoder for multiple code rates. For the low code rate $R{=}0.25$, the IB-IB \cite{shah_design_2019} decoder exhibits a loss of $0.2$\,dB w.r.t the double-precision LLR decoder while the proposed MS-CD decoder shows an approx. $0.08$ dB of additional degradation. Both the IB-IB and the MS-CD decoders for $R{=}0.25$ are constructed for a design $E_b/N_0{=}0.5$\,dB.  

The additional performance loss of MS-CD w.r.t the IB-IB decoder shrinks to approximately 0.05 dB at the code rate $R{=}0.5$. For the code rate $R{=}0.5$, Fig.~\ref{fig:BLER_4bit_multiple_rates} also includes the block error rate of MS-IB \cite{shah_MSIB_2023} decoder. The three finite alphabet decoders are constructed for design $E_b/N_0{=}0.5$\,dB. The error rate curve of the MS-IB decoder is in between the error rate curves of IB-IB and MS-CD decoders. This is expected behaviour since the MS-IB decoder design principle deploys an approximate, i.e., min-sum, design rule only for the upper branch while keeping the information bottleneck design rule for lower branch operations. The proposed MS-CD uses approximate design rules for both the upper and the lower branch operations.

The quantized decoders in Fig.~\ref{fig:BLER_4bit_multiple_rates} for the code rate $R{=}0.75$ are designed for $E_b/N_0{=}1.75$\,dB. It can be seen that the performance degradation shown by the MS-CD decoder w.r.t the IB-IB decoder reduces further at this high code rate. Similar trends have been noticed for LDPC decoders\cite{mohr_uniform_2022}.

\begin{figure}[tb]
	\centering
	\begin{tikzpicture}
\pgfplotsset{legend style={font=\small}}
\begin{axis}[%
width=0.8*\columnwidth,
height=0.67*\columnwidth,
scale only axis,
xmin=0.0,
xmax=4.9,
ymode=log,
ymin=8e-04,
ymax=1,
yminorticks=true,
ylabel={Block Error Rate},
xlabel={$E_b/N_0$ in dB},
axis background/.style={fill=white},
xmajorgrids,
ymajorgrids,
yminorgrids,
xminorticks=true, minor x tick num=4,
legend style={at={(0.01,0.01)}, anchor=south west},
legend style={legend cell align=left, align=left, draw=white!15!black}
]
\addplot [color=black,solid,mark=.,line width=1.0pt]
table{%
0					0.96552
0.25					0.86318
0.5					0.70268
0.75					0.43804
1					0.20766
1.25					0.071776
1.5					1.76E-02
1.75					3.12E-03
2					3.78E-04
2.25					3.38E-05
2.5					2.75E-06
	
};
\addlegendentry{LLR};%
\addplot [color=darkblue,solid,mark=o,line width=1.0pt]
table[row sep=crcr]{%
0	0.980456026058632\\
0.25	0.895833333333333\\
0.5	0.737745098039216\\
0.75	0.506734006734007\\
1	0.236263736263736\\
1.25	0.103011635865845\\
1.5	0.0342356687898089\\
1.75	0.00685867930547327\\
2	0.00113248152663025\\
2.25	0.000183\\
};
\addlegendentry{IB-IB};%

\addplot [color=darkred,solid,mark=x,line width=1.0pt]
table{%
0	0.974675108297234
0.250000000000000	0.912362545818061
0.500000000000000	0.773408863712096
0.750000000000000	0.540653115628124
1	0.297734088637121
1.25000000000000	0.129123625458181
1.50000000000000	0.0414631091068521
1.75000000000000	0.00928630730912334
2	0.00170512610593237
2.25000000000000	0.000277261355550965
};
\addlegendentry{MS-CD};%

\addplot [color=darkblue,solid,mark=o,line width=1.0pt]
table{%
0	0.995668110629790
0.250000000000000	0.973675441519494
0.500000000000000	0.905948017327558
0.750000000000000	0.767077640786405
1	0.550399866711096
1.25000000000000	0.321809396867711
1.50000000000000	0.144951682772409
1.75000000000000	0.0528052805280528
2	0.0141164372356089
2.25000000000000	0.00406195148523508
2.50000000000000	0.00101593538381837
2.75000000000000	0.000315059664032400
};

\addplot [color=darkred,solid,mark=x,line width=1.0pt]
table{%
0	0.995834721759414
0.250000000000000	0.976757747417528
0.500000000000000	0.922692435854715
0.750000000000000	0.792069310229923
1	0.595634788403865
1.25000000000000	0.358047317560813
1.50000000000000	0.175774741752749
1.75000000000000	0.0621560877315353
2	0.0199145710163786
2.25000000000000	0.00600974876859906
2.50000000000000	0.00172819685733293
2.75000000000000	0.000607129459145966
};

\addplot [color=darkblue,solid,mark=o,line width=1.0pt]
table{%
 0	1
0.250000000000000	1
0.500000000000000	0.999800066644452
0.750000000000000	0.999066977674109
1	0.994201932689104
1.25000000000000	0.976541152949017
1.50000000000000	0.920493168943686
1.75000000000000	0.812795734755082
2	0.631722759080307
2.25000000000000	0.428390536487837
2.50000000000000	0.254048650449850
2.75000000000000	0.135488170609797
3	0.0647912282644811
3.25000000000000	0.0303507948219196
3.50000000000000	0.0141985283735233
3.75000000000000	0.00673618709480532
4	0.00339810382052004
4.25000000000000	0.00161948970863501
4.50000000000000	0.000870078528192467
4.75000000000000	0.000402275044608718 
};

\addplot [color=darkred,solid,mark=x,line width=1.0pt]
table{%
0	1
0.250000000000000	1
0.500000000000000	0.999944462956792
0.750000000000000	0.999055870265467
1	0.993779851160724
1.25000000000000	0.975563700988559
1.50000000000000	0.914695101632789
1.75000000000000	0.801954903920915
2	0.631789403532156
2.25000000000000	0.425691436187937
2.50000000000000	0.251860490947462
2.75000000000000	0.136719791055914
3	0.0674246210228219
3.25000000000000	0.0306444449638858
3.50000000000000	0.0157567714302950
3.75000000000000	0.00770906257562968
4	0.00371965321211707
4.25000000000000	0.00194741095739256
4.50000000000000	0.00107736841627668
4.75000000000000	0.000564035523009065
};

\draw (axis cs:1.56,1.2e-2)  ellipse (0.1cm and 0.1cm);
\node at (axis cs:1.12,1.2e-2){64-bit};

\draw (axis cs:1.6,2e-2) ellipse (0.17cm and 0.17cm);
\node at (axis cs:1.15,2e-2){4-bit};

\draw (axis cs:2.0,2e-2) ellipse (0.3cm and 0.2cm);
\node at (axis cs:2.5,2e-2){3-bit};

\draw (axis cs:3.4,2e-2) ellipse (0.3cm and 0.2cm);
\node at (axis cs:3.9,2e-2){2-bit};
\end{axis}
\end{tikzpicture}%
	\caption{Block error rate comparison under CRC-aided SCL decoding with different message resolutions.}
	\label{fig:BLER_HalfRate_multiple_resolutions}
 \vspace{-0.5cm}
\end{figure}
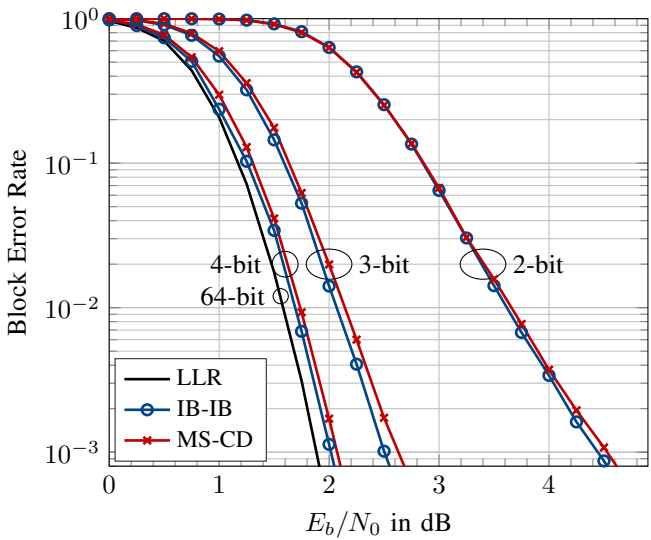
Fig. \ref{fig:BLER_HalfRate_multiple_resolutions} compares block error rates of the 4-bit IB-IB and MS-CD decoders of Fig. \ref{fig:BLER_4bit_multiple_rates} at code rate $0.5$ with their respective 3 and 2 bit variants. The 3-bit IB-IB and MS-CD decoders are designed for $E_b/N_0{=}0.5$\,dB while the 2-bit decoders are designed for $E_b/N_0{=}3.0$\,dB. It can be seen that by decreasing the decoder resolution from 4 to 3 bits, the gap between the IB-IB and MS-CD widens to 0.59\,dB. 
Varying resolutions within a decoder and extended design techniques could reduce the observed degradation under coarse quantization as shown in \cite{mohr_variable_2022} for LDPC decoding.

Another observation is the difference in the performance under the SC and CRC-aided SCL decoding of the IB-IB and MS-CD decoders constructed for the same design $E_b/N_0$. For $w{=}4$ bits, there is no difference in the error correction performance of IB-IB and MS-CD decoders as seen in Fig.~\ref{fig:BLER_SC}. However, a small difference can be seen in Fig.~\ref{fig:BLER_4bit_multiple_rates} when the same decoder is used for SCL decoding. 
It is not completely clear as what leads to this performance difference between the IB-IB and MS-CD under SC and SCL decoding. A major reason could be the fact that the decoders are constructed using quantized density evolution that assumes successive cancellation decoding. In other words, the decoder design framework is not aware of the list and the outer CRC used in the SCL decoding.

\section{Conclusions}
In this paper, finite alphabet decoders are designed for polar codes.
This class of decoders replaces LLR-based computations with mutual information maximizing table lookup operations. 
The main contribution is the use of a computational domain with uniform quantization instead of a lookup table for a significant complexity reduction in the lower branch update. In the case of 4-bit message resolution, we estimate only 1/25 of the memory consumption compared to a pure lookup table implementation. The uniform quantization requires only 1/4 of the computational cost compared to the optimal non-uniform quantization. The min-sum operation is chosen for the upper branch processing. It is shown that at 4-bit resolution, the performance degradation due to the used hardware-friendly approximations remains below 0.08 dB compared to the information-optimal lookup table design.%

\bibliographystyle{IEEEtran}
\bibliography{literature}
\end{document}